\documentclass{nrc1}

\usepackage{graphicx}
\usepackage{amsmath}
\usepackage{amssymb}
\usepackage{bm}
\usepackage{subfig}

\usepackage[english]{babel}

\title{Experimental High Energy Physics Summer School For High Schools}

\author{S. G\"{u}rb\"{u}z}
\address{Bo\u{g}azi\c{c}i University, Department of Physics, \.{I}stanbul, Turkey}
\correspond{saime.gurbuz@cern.ch}

\author{A. Ad{\i}g\"{u}zel}
\address{\.{I}stanbul University, Department of Physics, \.{I}stanbul, Turkey}

\author{V. E. \"{O}zcan}
\address{Bo\u{g}azi\c{c}i University, Department of Physics, \.{I}stanbul, Turkey}
\IDnote{Also at Feza G\"ursey Center for Physics \& Mathematics, Bo\u{g}azi\c{c}i University, \.{I}stanbul, Turkey}

\author{S. M. K{\i}rp{\i}c{\i}}
\address{Bo\u{g}azi\c{c}i University, Department of Physics, \.{I}stanbul, Turkey}

\author{A. Y{\i}lmaz}
\address{\.{I}stanbul Technical University, Department of Physics, \.{I}stanbul, Turkey}

\shortauthor{Gurbuz et al.}

\begin{document}

\maketitle

\begin{abstract}
Experimental High Energy Physics Summer School for High Schools, (Liseler \.{I}\c{c}in Deneysel Y\"{u}ksek Enerji Fizi\u{g}i Yaz Okulu - lidyef2018) was held between 9-16 September 2018 at Bo\u{g}azi\c{c}i University, Turkey, with financial support from T\"{U}B\.{I}TAK under the 4004 grant 118B491. Out of nearly 700 (11th and 12th grade) applicants, 30 had been selected from all around Turkey. Students were introduced to the fundamentals of high energy physics and performed experiments that demonstrated the techniques of this field, such as a salad-bowl electrostatic accelerator, and a cloud chamber. Here we report on the planning, implementation and the outcomes of the school that can serve as a template for similar activities in the future.


\end{abstract}

\section{Introduction}

While there is a substantial rise in implementing STEM \cite{stem12} applications into science education on the high school level \cite{stem15}, there still remains an inadequacy in the percentage of real-life applications in the science curriculum, especially in countries like Turkey. From a broader perspective, this leads to a conclusion that students' perceptions of science are not sufficiently associated with the measurable, testable and reproducible physical processes but rather with the applications of memorized mathematical expressions \cite{fen09}. The main underlying causes of students' misconceptions in the science applications can be attributed to the inadequacy or even non-existence of laboratory infrastructures, the orientation of the experimental setups towards the demonstration of mostly classical mechanical concepts, and the failure of such demonstrations in piquing the curiosity and/or enthusiasm of the students. These points, of course, surface if we can leave aside the general issues that affect the high school education on a more general level such as the large student population density and/or the shortage of qualified teachers, the inconsistency within the goals and objectives in the implementation of the general curriculum into the classrooms \cite{ozden2007}.

Furthermore, it has been argued that scientific literacy is best taught by
seeing science education as `education through science\cite{holbrook2009}.  However, experiments
in high school science courses are often detached from scientific frontiers
that excite many students, discoveries like the observation of gravitational waves
or the Higgs Boson.  Therefore, it is of
interest to design hands-on experiments that can be constructed and run by high school
students themselves, suitable for their experience and attentiveness
levels, yet still be connected to frontier fields such as cosmology and particle
physics. Towards that goal, we have attempted (a) to develop innovative high-school-level
experimental setups and documents that are connected to particle physics, (b) to test
the developed materials first with interns, and then (c) to convert these into a
week-long summer school program.

Our attempt was carried in the form of an experimental particle physics school held
in the summer of 2018 with financial support from T\"{U}B\.{I}TAK under the 4004 grant 118B491. For this organization: (1) An experienced team was formed from people who had prepared setups for CERN's high school contests, and/or supervised high school students, and/or provided training to Turkish high-school teachers for years at CERN. Actual researchers from CERN were also included in the team, as an extra means of improving the enthusiasm of the students. (2) Experimental setups were specifically designed to keep the technical and theoretical information required for comprehending the underlying processes at a minimum level for high school students. Considering the fact that not all of the students have the same scientific background, necessary accommodation was acheived by introducing lectures focusing on the basics. (3) The context of the experiments and the needed manual skills were selected from a wide range of possibilities in order to generate a wider range of opportunities for each of the students to enjoy and improve themselves. (4) Certain parts of the setups were chosen to allow participants to share their experiences with other students afterwards, and even perform entirely new experiments themselves.

In this proceeding, we report the application process, the student profile, the program and the outcomes of the school. Activities held, experiments performed and lectures given are summarized. Finally, we briefly describe the assessments and the evaluations performed during and after the school.

\section{Application Process}

Following the announcement of the school over social media, the applications were accepted over a period of about two weeks. The applicants were asked to have one reference letter submitted and were expected to fill in an online form, in which they provided (i) basic identification data (name, gender, address, name and location of the high school they are attending, grade), (ii) information on any relevant technical experience (Arduino, Raspberry Pi, 3D printers, and programming in general), (iii) a brief description of past scientific activities (school projects, attendance at science fairs, participation in summer schools, etc.), and (iv) the average grade points from maths and physics courses of the most recent semester. Finally they were asked a couple of open-ended qyestions like \textit{"What does science mean to you?"} and \textit{"Write down 3 questions you wish to find answers to when you attend lidyef."}.



\begin{figure}[hbt!]
\centering
\topcaption{\label{fig:application_geo} The poster and the geographic distribution of the applications}
\subfloat[The poster lidyef2018]{\includegraphics[height=0.3\textwidth]{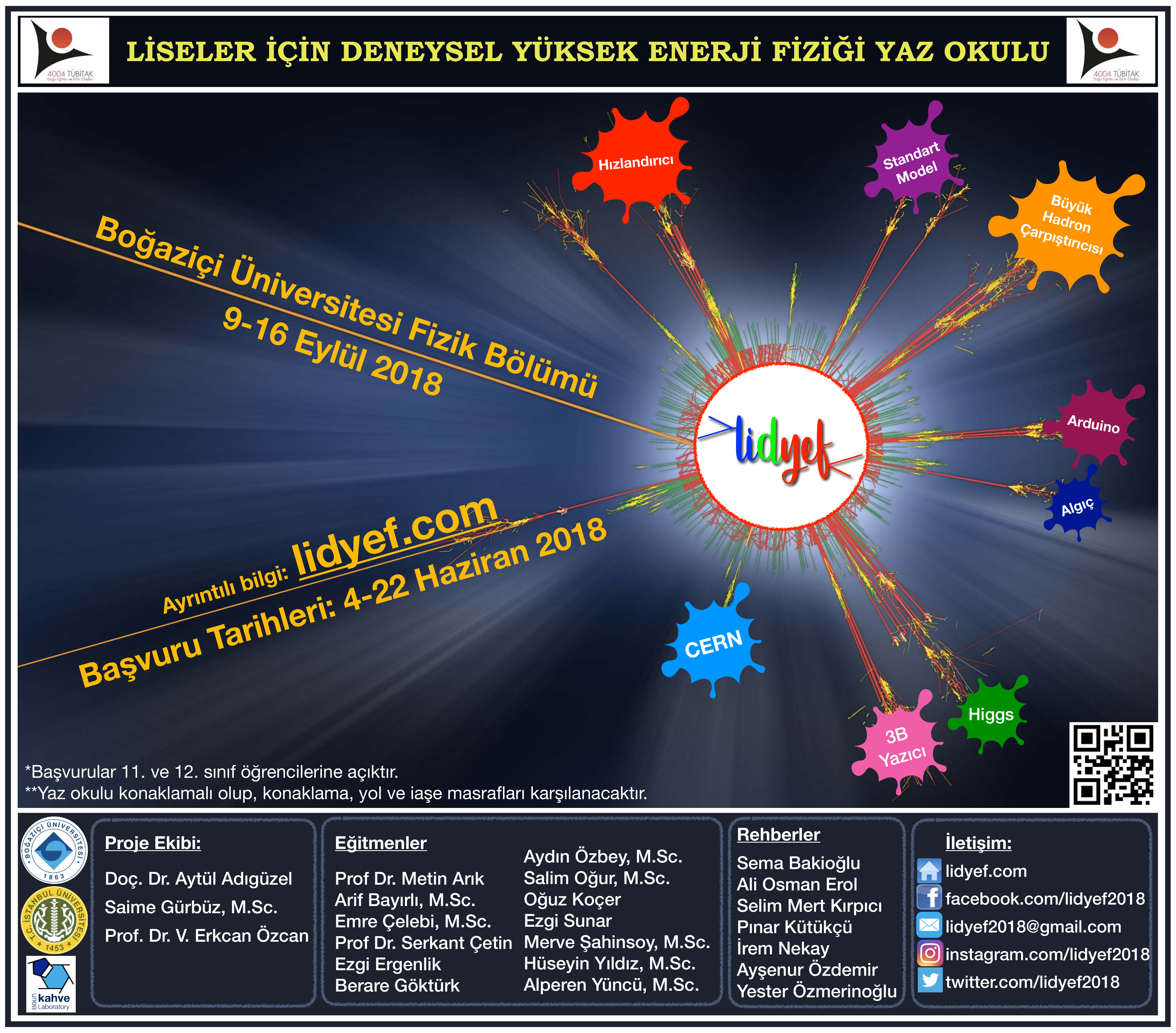} } \qquad \subfloat[The geographic distribution of the applications to lidyef2018]{\includegraphics[height=0.3\textwidth]{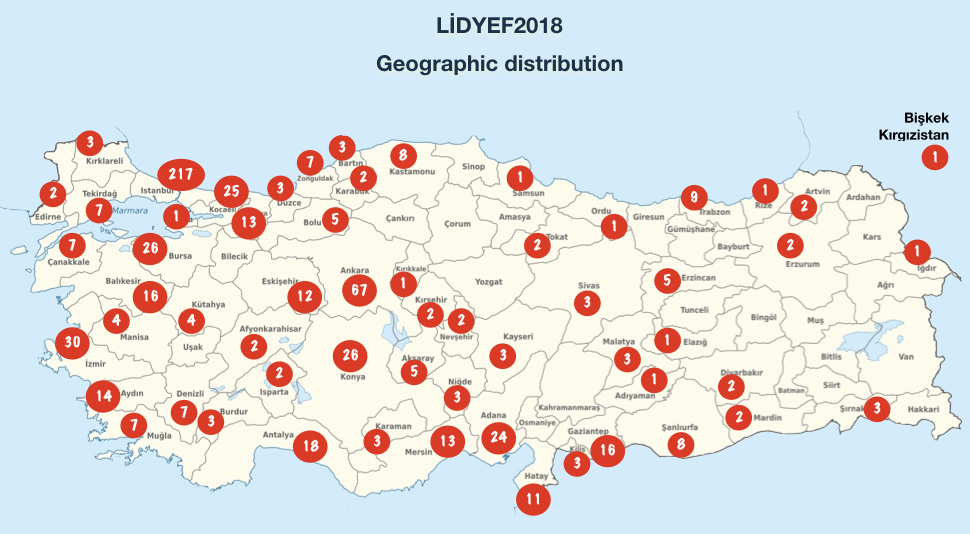} } 
\end{figure}

The school had initially been conceived with only 11th grade students in mind, but before the start of the application process a decision was made to accommodate a small quota of 12th graders in order to facilitate peer education and to evaluate the interest level of students who would soon start preparing intensively for the university entrance exam in Turkey. In total, 681 valid applications were received. Some statistics are provided below:

\begin{flitemize} 
\item \textbf{Gender distribution:} 44.5\% female, 55.5\% male.
\item \textbf{Grade distribution:} 52.1\% 11th grade, 47.9\% 12th grade.
\item \textbf{Distribution by province} is shown in Figure~\ref{fig:application_geo}.
\item \textbf{Type of school:} 41.9\% Anatolian high school, 30.7\% science high school, 9.1\% private Anatolian high school, 5.9\% private science high school, 5.1\% religious high school, 7.3\% other types.
\item \textbf{Last available physics grade:} 	88.4 $\pm$ 33.3 and 
\textbf{mathematics grade:} 	90.7 $\pm$ 13.3.
\end{flitemize}

A group of 4 academicians from the project team evaluated the applications. 30 students (24 from the 11th grade and 6 from the 12th grade) were selected, mostly based on their answers to the open-ended questions.  The aim of the open-ended questions was to gauge the level of their motivations and their perceptions of science.  Numerical measures such as the physics and math grades functioned only to eliminate the few students with insufficient technical background. To promote equality in opportunity, the students who had not had past opportunities to participate in science events were given preference. Although technical experience was not used as a selection criteria, a mix of experienced and inexperienced-but-highly-motivated students were aimed at the last step of the selection process. Finally, effort was spent to fairly match the fractions of students of a given gender (16 female, 14 male) and geographic location to the those of the national population.

\section{Teaching Techniques}

The school brought students together from different backgrounds with various abilities and personalities. In order to meet a broad spectrum of individual needs, we focused on implementing various student-focused teaching techniques.

In order to facilitate a better grasp of the real-world applications of the topics covered in the lectures, a significant amount of visualization was integrated in the descriptions of the concepts, and the descriptions were enriched by adding daily-life examples. The experiments used in the program were specifically designed to increase the inclusion of the students to the inquiry process by introducing semi-free hands-on activities instead of fully-guided cookbook-type experiments.

Throughout the program, the students were encouraged to work together in small groups (5 students in each group). By doing so, we aimed to engage the students in a cooperative learning \cite{coop16} process in which they were expected to work as a group with other students of different abilities.  Hence, they had the chance to experience a peer-oriented environment in which they could freely express their ideas and respond to each other, and could develop and/or improve their self-confidence while attaining the necessary communication and critical thinking skills. 

As a part of the program, we also implemented the inquiry-based teaching method \cite{dos15} by requiring students to work on projects of their own choice. Some basic guidelines for safety and originality of the work were established and supplies were obtained and provided to the students as needed. The students conceived and implemented their projects entirely by themselves (some individually, others in groups of 2-4) during their free times (mostly evenings at their dormitory).  Towards the end of the program, they were asked to present their work at an evening event, which stimulated lively discussions with the lecturers, project leaders and guide teachers.

Throughout the program, we focused on helping the students explore their own ideas and improve their problem-solving skills. In order to achieve this, in all of lectures and experiments, we prioritized a chain of thought-provoking questions as a source of inspiration for them to be able to have the thinking process on their own and become more independent as learners.

In order to accommodate the accelerated growth of technological improvements and to demonstrate the ubiquitous use of computers in particle physics, introductory-level lectures were included on the basics of programming and Arduino prototyping boards, and a Geiger counter application was implemented with Arduino.  

A disciplined yet friendly atmosphere of mutual respect was created for both the teachers and students.  This was facilitated by having the guide teachers to stay at the same lodging as the students.  Finally, after successful presentations of their projects, certificates of attendance and Arduino starter sets were handed out to the students, to award their contributions and to give them a chance to keep on exploring after they return to their high schools.

\section{Structure of the Program}

Lidyef-2018 program spanned a full week. Theory lectures were held in the mornings and experiments and applications in the afternoons. The students were expected to develop their own particle-physics-relate projects in the evenings to be presented at the end of the school.


\subsection{Meeting and Introduction}

On the first day, the students were picked up from the airports and bus terminals by the guide teachers. Once all had arrived, the program was introduced and the safety issues were explained by the project leaders and the project nurse. A small game was played to introduce students to one another.

\subsection{Theoretical Lectures}

Theory lectures were held in the mornings in two 40-minute sessions with a 10-minute break in between. The aim was to provide the theoretical background and prepare the students for the experiments and applications. The lectures were taught by experts (recent physics BSc graduates to full professors of particle physics). A complete list of lectures is provided below:

\begin{flitemize} 
\item \textbf{Modern Physics and Cosmic Particles:} Basic concepts of quantum physics and special relativity and cosmic particle physics with a historical context. 
\item \textbf{Particle Physics:} Review of the Standard Model and the elementary particles.
\item \textbf{Electricity and Magnetism:} Theory of electricity and magnetism for detector and accelerator physics. 
\item \textbf{About CERN:} Introduction to the laboratory, the Large Hadron Collider and its detectors.
\item \textbf{Detector Physics:} Short history, basic working principles and types of particle detectors. 
\item \textbf{Basic Analysis Methods:} Significant figures, experimental uncertainties, precision and accuracy. 
\item \textbf{Accelerator Physics:} Short history, basic working principles and types of particle accelerators. 
\item \textbf{Theoretical Particle Physics:} Overview of theoretical particle physics concepts; historical and conceptual construction of modern physics, progress from Newtonian mechanics towards quantum field theories.
\item \textbf{Applications of Particle Physics:} Applications in areas like medicine, computing, industry, etc. An engineering point of view into the world of particle physics. 
\end{flitemize}

\subsection{Computer Based Lectures and Applications}

\begin{figure}[hbt!]
\centering
\topcaption{\label{fig:computer} Computer based lectures and applications}

\subfloat[Arduino applications]{{\includegraphics[width=0.28\textwidth]{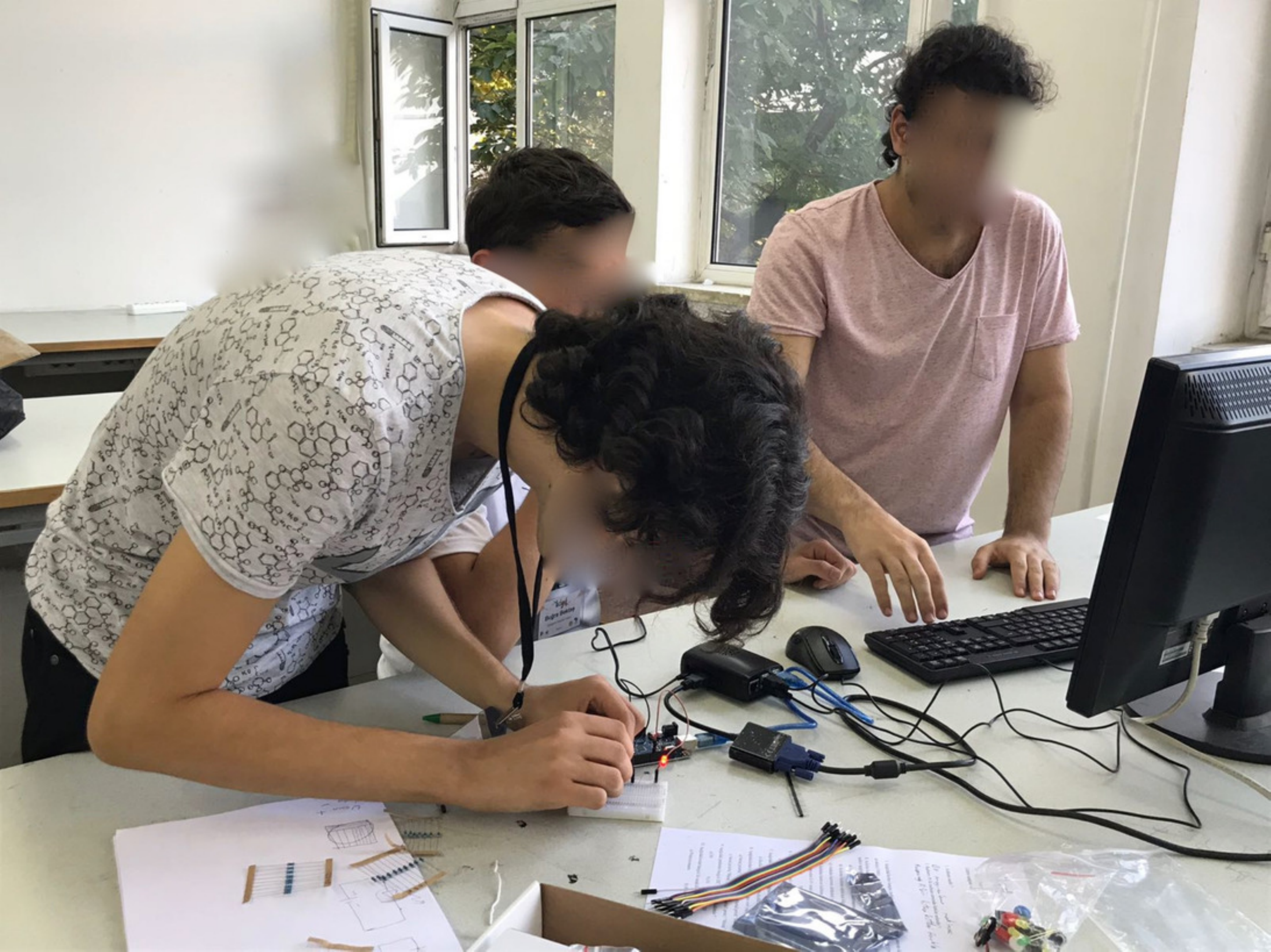} }} \qquad \subfloat[Geiger counter]{{\includegraphics[width=0.28\textwidth]{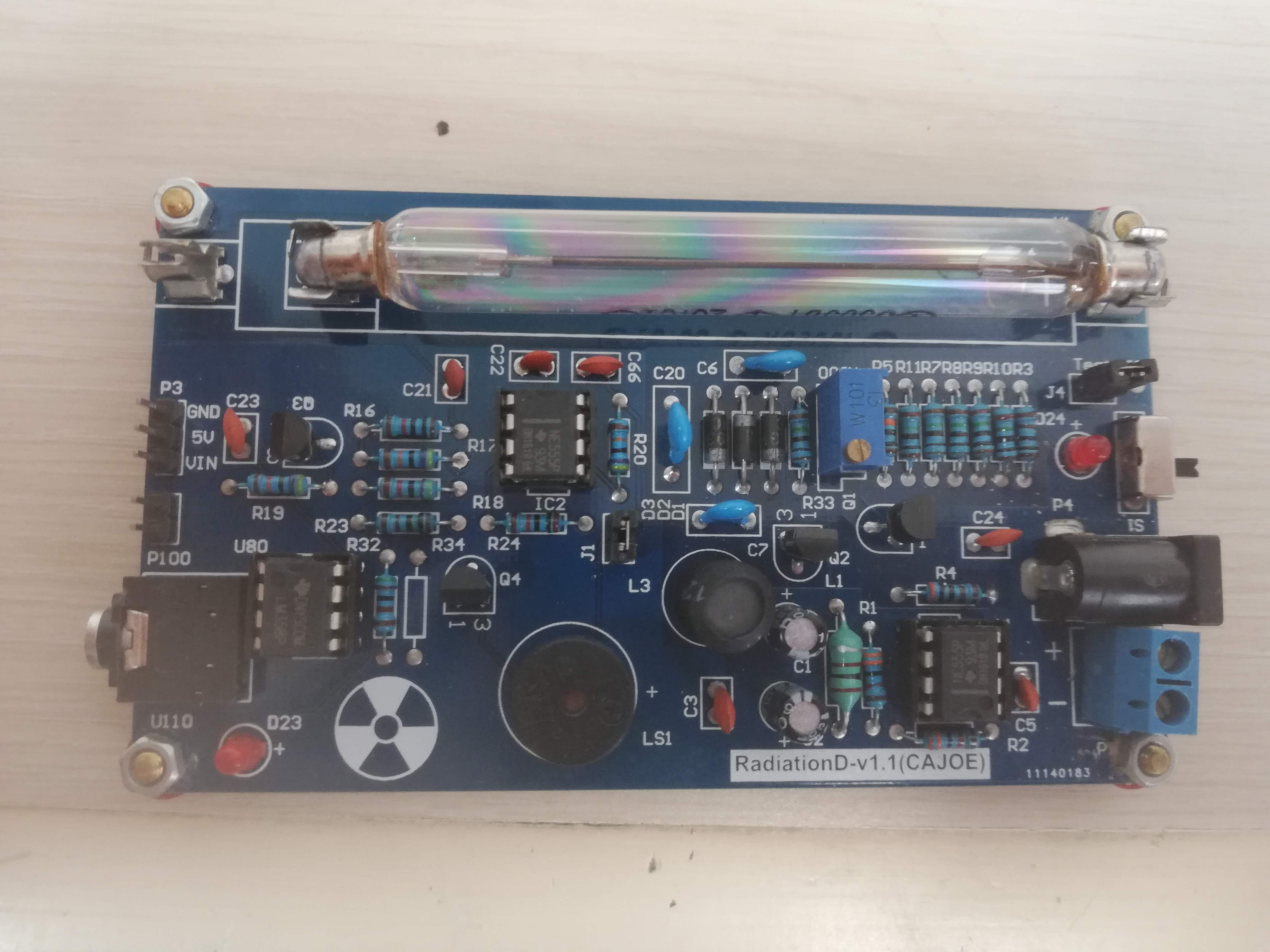} }} \qquad \subfloat[Hypatia screenshot]{{\includegraphics[width=0.28\textwidth]{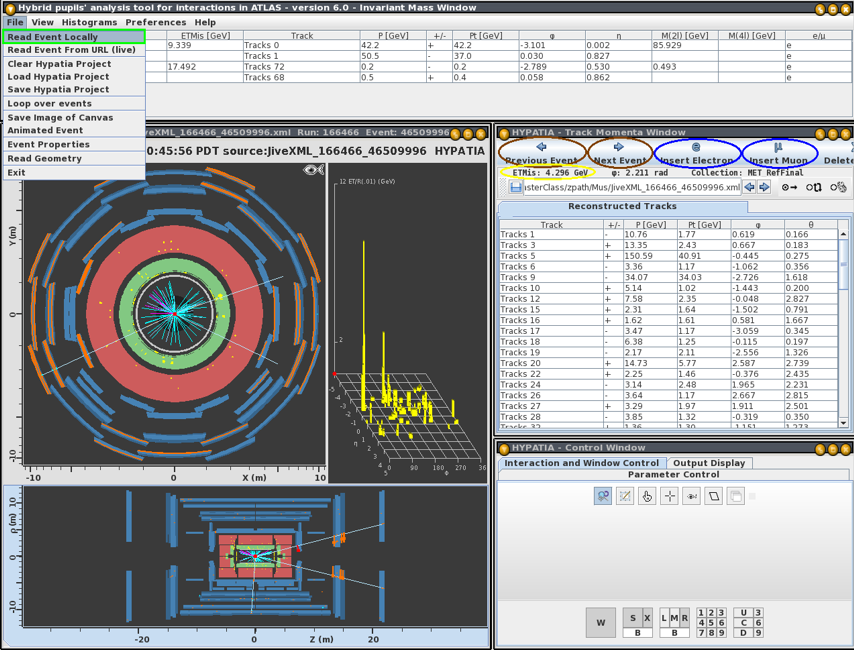} }} %
\end{figure}

A number of computer based lectures and application sessions were also included
in the program.  While they had initially been planned to span 90-minute periods,
upon the feedback received from the students, it was concluded that the students
would benefit more from longer sessions with longer discussion parts.  Hence the
duration of these lectures should be re-evaluated for future programs. The students
were split into groups of three during the application sessions
(Figure~\ref{fig:computer}). At the end of each application, they were given
report sheets to fill out.

\begin{flitemize} 
\item \textbf{Introduction to programming:} Introduce how computers work and the main principles and basic methods of computer programming. At the end of the lecture students were advised to play the online "light bot" game (http://lightbot.com/hour-of-code.html).
\item \textbf{Arduino lectures and applications:} Programming basic tasks with the Arduino IDE and introduction to taking data from sensors. In the hands-on session, the students were given LEDs, resistors, sensors, etc. and were expected to complete small sections of an already prepared source code that lights up the LEDs in a given pattern, and to print on the screen digital and analog data read from the sensors. 
\item \textbf{Geiger counter with Arduino:} A Raspberry Pi 3+, an Arduino Uno and a Geiger counter were provided to the students, as well as source code that prints the time that a particle passes through the counter. They were expected to take 6 minutes of data and draw histograms of the counts in 30-sec and and 1-min bins and comment on what they have seen.
\item \textbf{Hands-on CERN ATLAS experiment data:} CERN has been supporting so-called Masterclass events for years where high school students analyze data from actual collision events collected by the ATLAS or CMS experiments. At lidyef2018, we followed the $Z$-path of the ATLAS Masterclass \cite{masterclass2018}. The students were introduced to the ATLAS Detector geometry, event reconstruction and software. Then they were expected to analyze $Z\rightarrow\ell\ell$ events using HYPATIA software \cite{hypatia} and reconstruct the mass of the $Z$ boson.

\end{flitemize}

\subsection{Experiments}

\begin{figure}[hbt!]
\centering
\topcaption{\label{fig:experiments} Experiments}

\subfloat[Cloud chamber setup]{\includegraphics[height=0.15\textwidth]{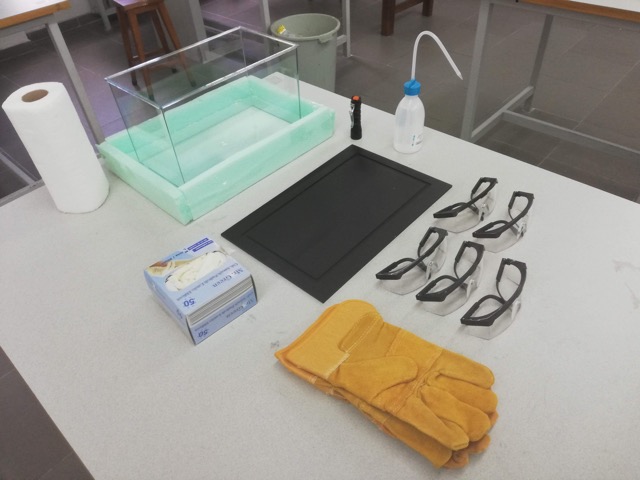} } 
\subfloat[Laser diffraction experiment]{\includegraphics[height=0.15\textwidth]{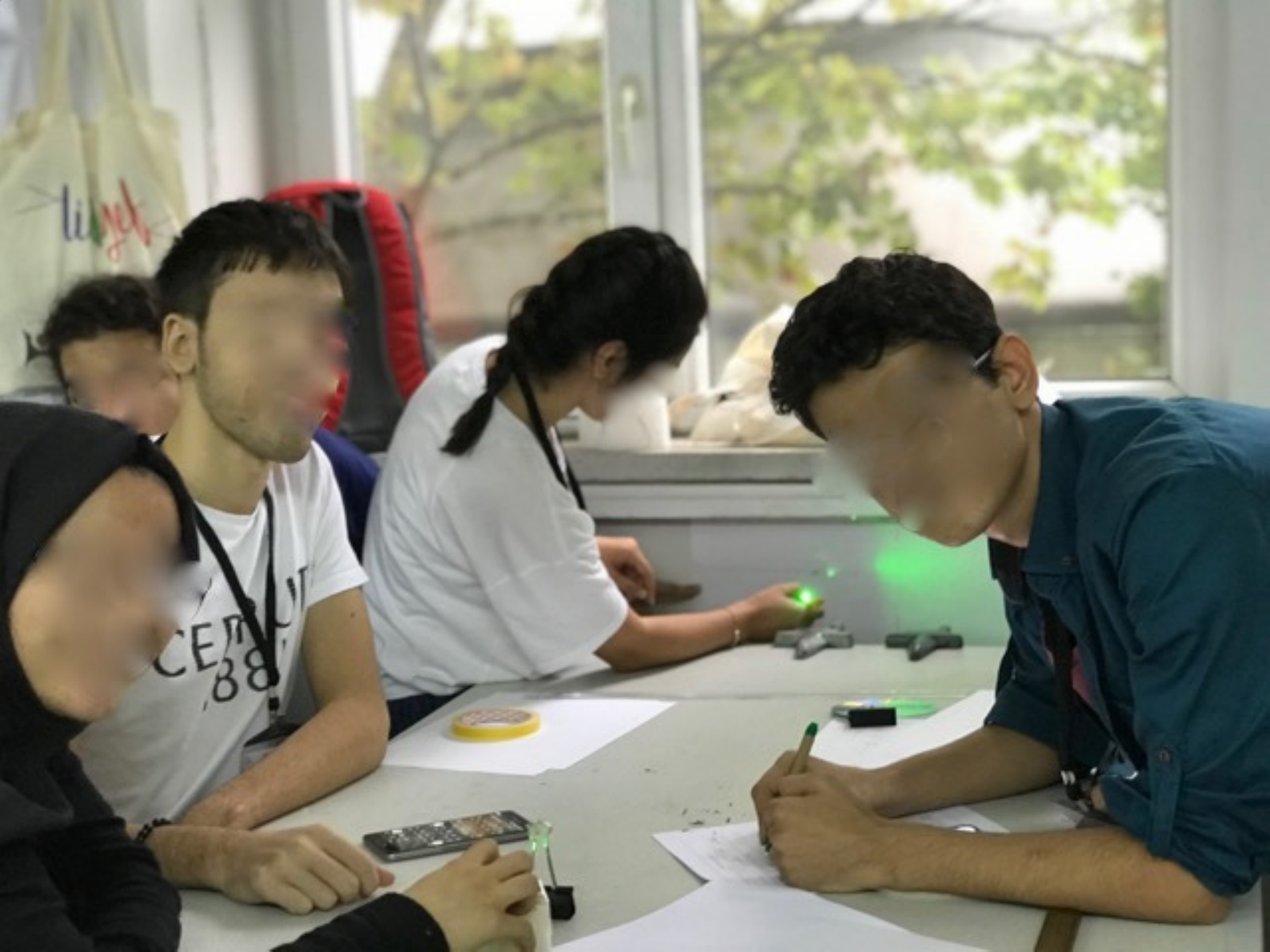} } 
\subfloat[Interns measuring the speed of light with chocolate on microwave]{\includegraphics[height=0.15\textwidth]{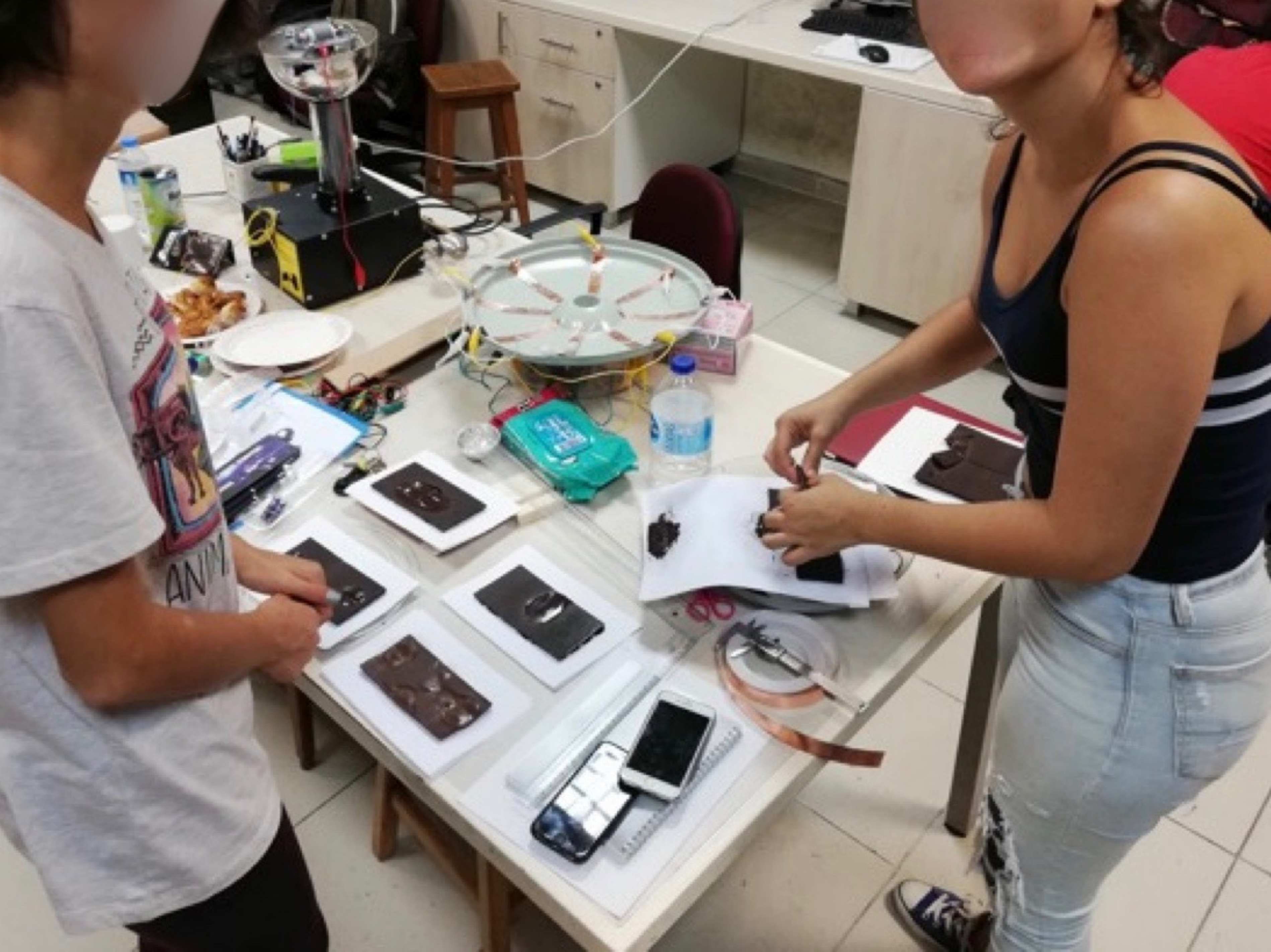} }  \subfloat[Salad bowl experiment]{\includegraphics[height=0.15\textwidth]{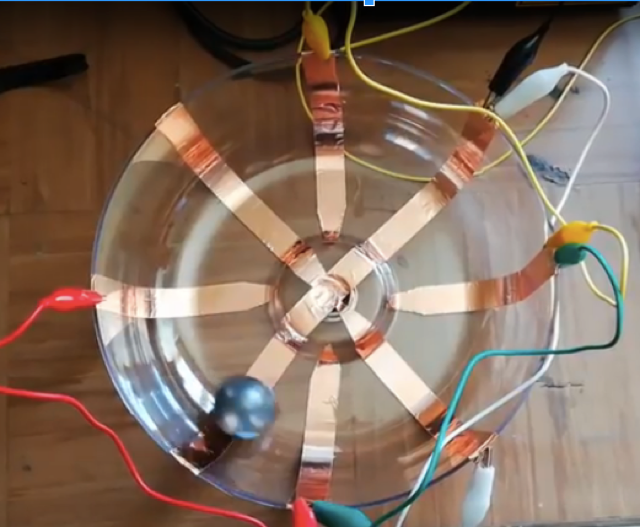} } 
\subfloat[Model of the ATLAS toroid magnet]{\includegraphics[height=0.15\textwidth]{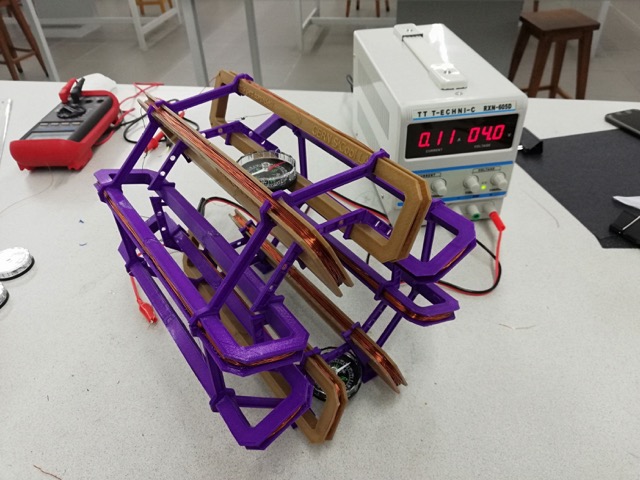} } %
\end{figure}

Given the budget constraints six copies of each setup was prepared, and the students were split into groups of five to run the experiments concurrently.  Before the start of the school, all the setups had been tested by two summer interns, who were themselves high school students.  For each experiment, a report sheet was prepared to be filled by each group during the experiment and to be submitted at the end.  The reports included the following parts: aim of the experiment, materials used, observations/data collected.  The duration of each session was set to 90 minutes, but based on our observations, we would recommend extending this period to 2 hours in the future programs. The five different experiments that were carried out can be seen in Figure~\ref{fig:experiments}.

\begin{flitemize} 
\item \textbf{Wilson Cloud Chamber Experiment:} The cloud chamber is not only a detector which lead to the Nobel Prize winning discovery of positrons, and of muons and kaons, but is also used for educational purposes in particle physics. In the experiment, alcohol cloud is formed in a clear aquarium. In a dark room, with the help of a torch, the students could see the tracks of cosmic particles. At the end of the experiment, they discussed the qualitative differences between the observed tracks and which particles those tracks belong to. The background information provided to the students covered cosmic rays and the interactions of particles with matter.
\item \textbf{Diffraction Experiment:} To observe the diffraction of light, a common laser pointer, a CD or DVD, ruler and paper was used. Using data about the CDs and observing the interference patterns, first the frequency of red and green light from lasers were computed. Next, using the obtained frequency values, diffraction pattern from a single strand of hair was studied and its thickness was measured. The students were provided background information on various modern physics concepts, especially about light.
\item \textbf{Measuring the Speed of Light with Chocolate in a Microwave Oven:} Before this experiment, the students were provided background on the physics of waves and light. The turntable in the microwave owen was removed and two flat bars of chocolate (15.5$\times$7.5\,cm) were placed inside. The standing waves created in the microwave oven caused the chocolate to melt only at certain points: the nodes of the wave. By measuring the distance between the nodes, students obtained the wavelength and then calculated the speed of light. In Figure~\ref{fig:experiments} (c), the summer interns can be seen performing this experiment.
\item \textbf{Salad Bowl Experiment}: To demonstrate how electrostatic accelerators work, a salad bowl accelerator model was constructed. Eight strips of conductive (copper) bands were placed on a salad bowl, and they were charged with static electricity obtained from a Van de Graff generator. The connections were done in a way that caused neighbouring bands to be oppositely charged. A ping pong ball coated with a conductive paint (or painted with graphite from a pencil) was placed in the bowl. At each strip it collected alternating electric charges and moving from one strip to the next it got accelerated. The students calculated the speed of the ball in the accelerator and compared the model with accelerators like the Large Hadron Collider. Students were provided background information on electricity, magnetism and accelerator physics.
\item \textbf{ATLAS Toroid Model:} A fully working prototype of CERN's ATLAS detector's toroid magnet model can be built using a 3D Printer, copper wire and a low voltage power supply. The parts in the reference \cite{scoollab2016} were printed and glued. The coils were loaded with 80 turns of copper wire. Then all the parts were put together and connected to the power supply. The students observed the magnetic field lines using small compasses. After the experiment, a cathode ray tube was placed in the magnetic field of a pair of Helmholtz coils and the instructors demonstrated how electron trajectories are bent in a magnetic field. 
\end{flitemize}

\subsection{Visits and Live Connection to CERN}

The program included a number of \textit{extra curricular} visits, selected to complement the scientific program and also to provide a breathing space to the students. The destinations were: Sak{\i}p Sabanc{\i} Museum; \.{I}stanbul University Astronomy Department, Plenaterium, and Physics Department Laboratories; Bo\u{g}azi\c{c}i University South Campus, Physics Department, Kandilli Solar Observatory and Kandilli Detector, Accelerator and Instrumentation Lab (KahveLab). In addition to the visits, an hour-long live teleconference session was held, in which three Turkish scientists (a PhD student and two senior physicists) working at CERN introduced themselves and answered questions from the students. The aim was to inspire the students and give them a chance to meet scientists working at an international lab.

\section{Assessment and Evaluation}

Throughout the program, we implemented various methods in order to improve the validity and reliability of the assessment process of the school which are discussed in more detail below. 

\subsection{The Evaluation Survey}

As an assessment tool for the overall success of the program, we prepared an evaluation survey and distributed to the students at the last day of the school. The survey involved questions related to the evaluation of the school program, instructors, guides and experiments and applications in the Likert scale (out of 5, where 1 means \textit{``Very unsatisfied''} and  5 means \textit{``Very satisfied''}). To briefly summarize the results, students rated the program with a high overall score of 4.09 $\pm$ 0.77.  The content was found to be sufficient (4.27 $\pm$ 0.87), and the students stated that they would use their gains in the future (4.80 $\pm $ 0.61). They were very pleased with their instructors (4.69 $\pm $ 0.19), regarding them as experts on their fields (4.83 $\pm $ 0.38) and stated having good communication with them (4.80 $\pm $ 0.41).  Similarly, they found their communication with guides as favourable (4.81 $\pm $ 0.41) and all agreed that guides were always helpful and had lead them properly (4.77 $\pm$ 0.57). They also agreed that the experiments and lectures had appropriately been designed for their levels (4.40 $\pm $ 1.04 and 4.14 $\pm $ 0.45 respectively), test equipment were in a good shape (4.57 $\pm $ 0.73), and the documentation explaining the experiments were mostly clear (3.97 $\pm $ 1.00). Additionally, they were satisfied about the social program (4.19 $\pm $ 0.41). Median evaluations were usually 4 or 5. Lowest points (1-2) were rarely given and for a few questions.   We consider itself a positive sign that the students took the survey seriously, did not hesitate to criticize things they found to be insufficient and proposed improvements.

\subsection{Assessment and Evaluation of the Computing Applications}
A short test of 10 questions was issued to students in order to evaluate the comprehension of the computing lecture and its applications. The students scored an average of 6.62 out of 10, indicating that the lecture had met its basic objectives. At the end of the computing exercises, most of the students were observed to have written their own software using these Ardunio and Raspbery Pi cards in accordance with the objectives of the lecture.

\subsection{Discussion and Evaluation}
At the end of the school, a one-hour meeting was held in order to discuss and evaluate the performance. Below are some inferences and recommendations proposed by instructors, guides and students:
\begin{itemize}
\item All of the students agreed that schools with similar structure and curriculum should be organized regularly, and other students should be given this opportunity as well.
\item The project team and students agreed that networking among students and instructors was very important, and could be useful in the future.
\item It was proposed to organize the same program for high school teachers.
\item The students agreed that all project members were self-sacrificing and helpful during the school.
\item The students indicated that they had learned fundamentals of the inquiry process and felt highly motivated towards joining academia.
\end{itemize}

\subsection{Experiment reports} 
The student-filled reports from the six experiments were evaluated by two teachers of the program. Students scored an average of 4.2 out of 5. From this score, we have concluded positively about their ability for conducting experiments, writing reports, and preparing experimental setups. 


\subsection{Study of the Particle Physics Data}
As part of the ATLAS data analysis exercise, the students were expected to search for a track of $W$ particles by using the Minerva Software. Most of them achieved to identify 7 tracks out of 10. The fact that 3 tracks were missed was taken as a good indication that the time assigned for the task was not enough and should be increased for future applications.  We also delivered a test at the end of this exercise. The average score was 10.7 of out of 12. This score supports that the objective of the task, which was to impart information about particles, the ATLAS detector and basic analysis procedures, was achieved.

\subsection{Project Work} 
As a part of the program, students were expected to work on projects of their own during their free time. All the students participated enthusiastically, with a couple of students contributing to more than one project. A total of ten projects were presented at the end of the school: they had designed games, written books for children, and built lively detector demo boards with LEDs and Arduinos, all demonstrating or teaching the topics covered throughout the week. The presentations were also very colorful and the students were observed to be excited to showcase their products. The breadth and ingenuity of the projects also indicated that the students had been able to obtain the basic knowhow for accessing the necessary information, and for designing and developing products. 

\section{Conclusions}

The school was successfully held between 9-16 September 2018. The results of the assessment procedure discussed above show that both the students and the high school teachers considered the program to be immensely positive. A large fraction of the evaluation forms from the students indicated that the school had a huge impact on how they view the world and the role science plays in it, with many students expressing a desire to choose careers in STEM fields. The student projects were also found to be highly innovative, even by the high school teachers who are familiar with the education system in Turkey. 

Assessment procedures carried out throughout the week and feedback gathered during and after the lectures and experiments produced reliable results to conclude that the program did meet and surpass its objectives. To interested parties who want to organize similar events, we will make available the video recordings of the lectures, applications and experiments as well as the collected data from assessment methods.  Furthermore, we prepared guidelines that can allow secondary education institutions to implement similar experimental setups for their own students \cite{lidyef2018}.  We also foresee that the project will make a valueable contribution in increasing the success rate of students from Turkey when they participate in international contests organized by CERN or similar bodies.

\section*{Acknowledgements}
We wish to express our most sincere gratitude to those institutions and persons without whom lidyef2018 would not happen. We thank Cihan \c{C}i\c{c}ek and Assoc. Prof. Fatih Mercan for their great support while developing and submitting this project; Bo\u{g}azi\c{c}i University Department of Physics for providing us with the necessary lab spaces and classrooms; \.{I}stanbul University, TOBB ETU and KahveLab for their support; \.{I}stanbul Beyo\u{g}lu Anadolu High School for letting us use their 3D printers; KahveLab summer interns Do\u{g}a Aksen and Derin Sivrio\u{g}lu for their help in testing the experimental setups; our `guide' teachers Selma Erge, Ali Osman Erol, \.{I}rem Nekay,  Ay\c{s}enur \"{O}zdemir, Yester \"{O}zmerino\u{g}lu, Ahmet Renklio\u{g}lu, Reyhan \"{O}z Y{\i}ld{\i}z for their support throughout the entire school; our instructors Metin Ar{\i}k, Emre \c{C}elebi, Serkant \c{C}etin, Berare G\"{o}kt\"{u}rk, O\u{g}uz Ko\c{c}er, Salim O\u{g}ur, Ayd{\i}n \"{O}zbey, Sezen Sekmen, Ezgi Sunar, G\"{o}khan \"{U}nel, H\"{u}seyin Y{\i}ld{\i}z, Alperen Y\"{u}nc\"{u} for their valuable lectures; Bo\u{g}azi\c{c}i University undergraduate students Sevim A\c{c}{\i}ks\"{o}z and Ekin Nur Cang{\i}r for their voluntary help whenever necessary, our friends Ezgi Ergenlik, Y{\i}lmaz Ergenlik and Mustafa G\"{u}rb\"{u}z for their voluntary local help.

\end{document}